\documentclass[twocolumn,aps,superscriptaddress,prl]{revtex4}
\usepackage{graphicx}
\usepackage{xcolor}
\usepackage{times}
\usepackage{amsmath}
\usepackage{amssymb}
\usepackage{units}
\usepackage{stmaryrd} 
\DeclareGraphicsExtensions{.pdf,.png,.eps,.jpg}

\begin{document}
\preprint{0}

\title{Evidence for a Strong Topological Insulator Phase in $\mathbf{ZrTe_5}$}

\author{G. Manzoni} 
\affiliation{Universit\'a degli Studi di Trieste - Via A. Valerio 2, Trieste 34127, Italy} 

\author{L. Gragnaniello} \email[E-mail address: ]{luca.gragnaniello@uni-konstanz.de }
\affiliation{University of Konstanz, 78457 Konstanz, Germany} 

\author{G. Aut\`es}
\affiliation{Institute of Physics, Ecole Polytechnique F\'ed\'erale de Lausanne (EPFL), CH-1015 Lausanne, Switzerland}
\affiliation{National Center for Computational Design and Discovery of Novel Materials MARVEL, Ecole Polytechnique F\'ed\'erale de Lausanne (EPFL), CH-1015 Lausanne, Switzerland}

\author{T. Kuhn}
\affiliation{University of Konstanz, 78457 Konstanz, Germany} 

\author{A. Sterzi}
\affiliation{Universit\'a degli Studi di Trieste - Via A. Valerio 2, Trieste 34127, Italy} 

\author{F. Cilento} 
\affiliation{Elettra - Sincrotrone Trieste  S.C.p.A., Strada Statale 14, km 163.5 Trieste, Italy} 

\author{M. Zacchigna}
\affiliation{C.N.R. - I.O.M., Strada Statale 14, km 163.5 Trieste, Italy}

\author{V. Enenkel}
\affiliation{University of Konstanz, 78457 Konstanz, Germany} 

\author{I. Vobornik}
\affiliation{C.N.R. - I.O.M., Strada Statale 14, km 163.5 Trieste, Italy}

\author{L. Barba} 
\affiliation{CNR - Institute of Crystallography, Area Science Park, SS 14, km 163.5 Trieste, Italy} 

\author{F. Bisti}
\affiliation{Swiss Light Source, Paul Scherrer Institut, CH-5232 Villigen, Switzerland}

\author{Ph. Bugnon}
\affiliation{Institute of Physics, Ecole Polytechnique F\'ed\'erale de Lausanne (EPFL), CH-1015 Lausanne, Switzerland}

\author{A. Magrez}
\affiliation{Institute of Physics, Ecole Polytechnique F\'ed\'erale de Lausanne (EPFL), CH-1015 Lausanne, Switzerland}

\author{V. N. Strocov}
\affiliation{Swiss Light Source, Paul Scherrer Institut, CH-5232 Villigen, Switzerland}

\author{H. Berger}
\affiliation{Institute of Physics, Ecole Polytechnique F\'ed\'erale de Lausanne (EPFL), CH-1015 Lausanne, Switzerland}

\author{O. V. Yazyev}
\affiliation{Institute of Physics, Ecole Polytechnique F\'ed\'erale de Lausanne (EPFL), CH-1015 Lausanne, Switzerland}
\affiliation{National Center for Computational Design and Discovery of Novel Materials MARVEL, Ecole Polytechnique F\'ed\'erale de Lausanne (EPFL), CH-1015 Lausanne, Switzerland}

\author{M. Fonin}
\affiliation{University of Konstanz, 78457 Konstanz, Germany}

\author{F. Parmigiani}
\affiliation{Universit\'a degli Studi di Trieste - Via A. Valerio 2, Trieste 34127, Italy} 
\affiliation{Elettra - Sincrotrone Trieste  S.C.p.A., Strada Statale 14, km 163.5 Trieste, Italy} 
\affiliation{International Faculty - University of K\"oln, 50937 K\"oln, Germany} 

\author{A. Crepaldi} \email[E-mail address: ]{alberto.crepaldi@epfl.ch}
\affiliation{Institute of Physics, Ecole Polytechnique F\'ed\'erale de Lausanne (EPFL), CH-1015 Lausanne, Switzerland}
\affiliation{Elettra - Sincrotrone Trieste  S.C.p.A., Strada Statale 14, km 163.5 Trieste, Italy}

\date{\today}

\begin{abstract}

The complex electronic properties of $\mathrm{ZrTe_5}$ have recently stimulated in-depth investigations that assigned this material to either a topological insulator or a 3D Dirac semimetal phase. Here we report a comprehensive experimental and theoretical study of both electronic and structural properties  of $\mathrm{ZrTe_5}$, revealing that the bulk material is a strong topological insulator (STI). By means of angle-resolved photoelectron spectroscopy, we identify at the top of the valence band both a surface and a bulk state. The dispersion of these bands is well captured by \emph{ab initio} calculations for the STI case, for the specific interlayer distance measured in our x-ray diffraction study. Furthermore, these findings are supported by scanning tunneling spectroscopy revealing the metallic character of the sample surface,  thus confirming the strong topological nature of $\mathrm{ZrTe_5}$.

\end{abstract}

\maketitle

The discovery of topological insulators (TIs), characterized by metallic spin-polarized surface states connecting the bulk valence and conduction bands  \cite{Hasan_RMP_2010}, has stimulated the search for novel topological phases of matter \cite{XU_Science_2011, Sakano_natcom_2015, Liu_NatMat_2014, Xu_Science_2015, Gabriel_NatMat_16}.  $\mathrm{ZrTe_5}$ has recently emerged as a challenging system with unique, albeit poorly understood, electronic properties \cite{Skelton_1982, Jones_1982, Tritt_PRB_1999, Weng_PRX_2014, Valla_arx_2015, Zhou_arx_2015, Chen_PRL_2015, Chen_PRB_2015, Manzoni_PRL_2015, Niu_arxiv_2015, pariari_arxiv_16}. Magneto-transport \cite{Valla_arx_2015}, magneto-infrared \cite{Chen_PRL_2015} and optical spectroscopy \cite{Chen_PRB_2015} studies describe $\mathrm{ZrTe_5}$ in terms of a 3D Dirac semimetal. Theoretical calculations have predicted its bulk electronic properties to lie in proximity of a topological phase transition between a strong and a weak TI (STI and WTI, respectively), where only the former displays topologically protected surface states at the experimentally accessible (010) surface \cite{Weng_PRX_2014}. The monolayer is also computed to be a 2D TI \cite{Weng_PRX_2014} and scanning tunneling microscopy/spectroscopy (STM/STS) experiments suggest the existence of topologically protected states at step edges \cite{Li_arxiv_2016, Wu_arxiv_2016}. However, the unambiguous identification of the topological phase of $\mathrm{ZrTe_5}$ is still lacking.

In this Letter we report on the STI character of the bulk $\mathrm{ZrTe_5}$ by combining \emph{ab initio} calculations and multiple experimental techniques, at temperature both above and below the one of the resistivity peak, $\mathrm{T^*\,\sim}$  160 K  \cite{Skelton_1982, Jones_1982, Tritt_PRB_1999,Manzoni_PRL_2015}. Angle-resolved photoelectron spectroscopy (ARPES) experiments in the ultraviolet (UV) and soft x-ray (SX) energy ranges reveal the presence of two distinct states at the top of the valence band (VB). On the basis of photon energy dependent studies, we ascribe the origin of these two states to the bulk and crystal surface, respectively.  We have performed \emph{ab initio} calculations of the topological phase diagram of $\mathrm{ZrTe_5}$, as a function of the interlayer distance \emph{b}/2. Our measured band dispersion is in agreement with the calculations and it is consistent with the STI case for \emph{b}/2 = 7.23 $\mathrm{\pm}$ 0.02 $\mathrm{\AA}$. This value has been confirmed for our specimen by x-ray diffraction (XRD) measurements. Furthermore, the 3D Dirac semimetal phase is not protected by crystalline symmetries, and it manifests only for the specific \emph{b}/2 = 7.35  $\mathrm{\AA}$ at the boundary between the STI and the WTI phases. Finally, STM-STS experiments confirms the metallic character of the surface termination of the bulk $\mathrm{ZrTe_5}$. Altogether our results indicate that $\mathrm{ZrTe_5}$ is a STI both above and below $\mathrm{T^*}$, in proximity to the WTI phase. 


\begin{figure}[t!]
  \centering
   \includegraphics[width = 0.35 \textwidth]{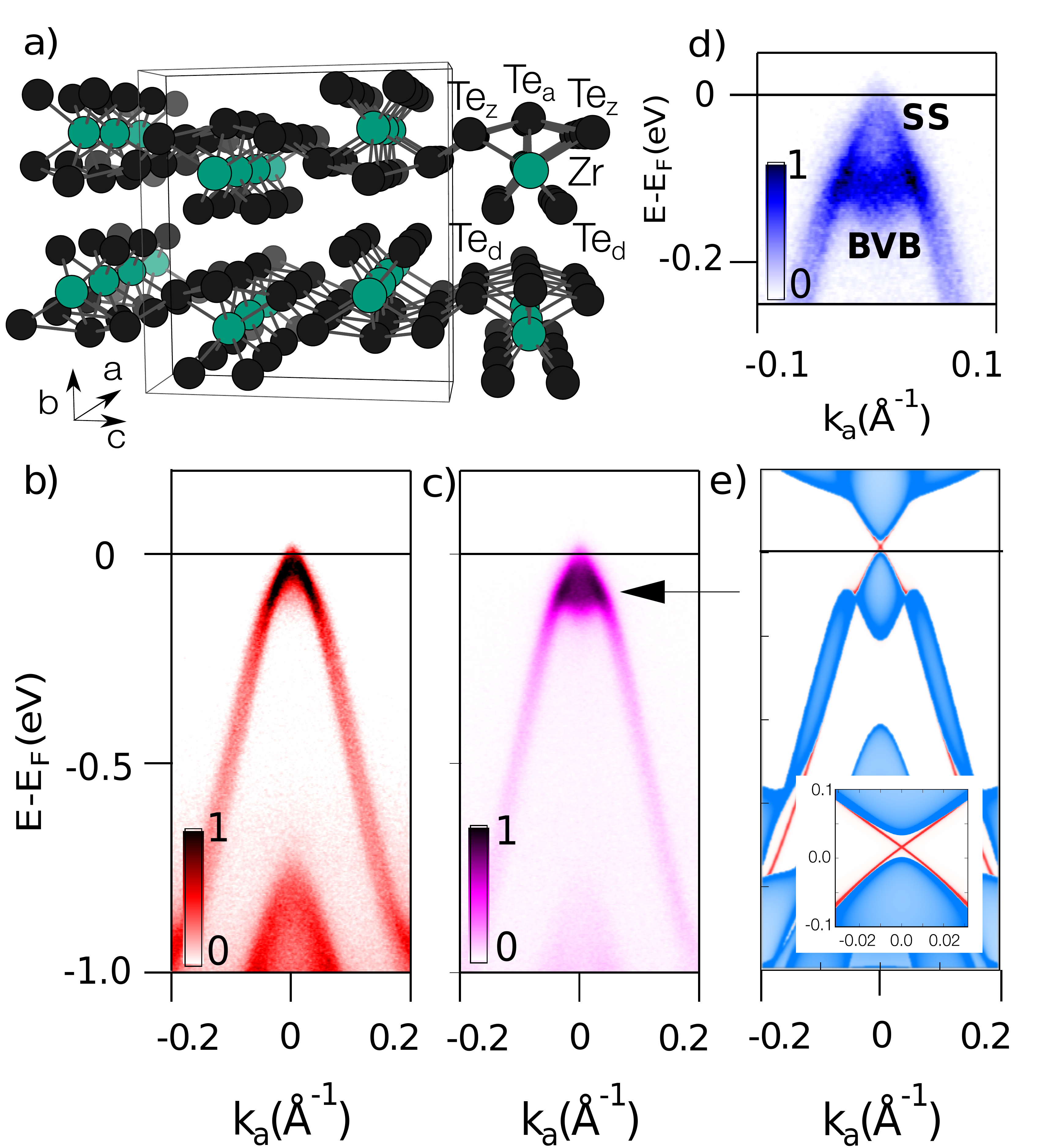}
  \caption[didascalia]{ (a)  Crystal structure of $\mathrm{ZrTe_5}$. (b) and (c) UV ARPES measurements along the chain direction $\mathrm{k_a}$, for 23.5 and 29 eV corresponding to $\mathrm{\Gamma X}$ and $\mathrm{Y X_1}$. At $\mathrm{Y} $ two states are observed, one crossing $\mathrm{E_F}$ and the other forming a \emph{M}-like shape reaching its maximum below $\mathrm{E_F}$.  (d) Zoom at $\mathrm{E_F}$ measured at 22 eV where the two states at the top of VB are more clearly resolved. (e) Calculated momentum resolved bulk density of states projected on the (010) surface (blue) and momentum resolved surface density of states at the (010) surface (red), for \emph{b}/2 =  7.23 $\mathrm{\AA}$. The surface state is more clearly visible in the zoom at $\mathrm{E_F}$ in the inset.  
 
   }
  \label{fig:ARPES_bands}
\end{figure}


High quality $\mathrm{ZrTe_5}$ single crystals have been grown by vapor transport technique \cite{Berger_1983}. The UV ARPES measurements have been carried out at the APE beamline, at the Elettra synchrotron, at $\sim$\,210\,K with linear horizontal (LH) polarization in the energy range 20-36 eV, with energy and angular resolution better than 20\,meV and 0.2\,$\mathrm{^{\circ}}$, respectively.  The SX ARPES experiments have been carried out at the ADDRESS beamline \cite{adress}, at the Swiss Light Source, Paul Scherrer Institute, Switzerland. The photon energy dependence has been investigated for both LH and vertical (LV) polarizations in the energy range 310-510 eV, with energy and angular resolution varying between 60 - 70\,meV and 0.07\,$\mathrm{^{\circ}}$, respectively. In this set of measurements the sample temperature was kept at $\sim$\,15 K in order to suppress Debye-Waller-like reduction of the coherent spectral weight \cite{Braun_PRB_13}. Single crystal x-ray diffraction (XRD) measurements have been performed for 8.85 keV photon energy at the Elettra storage ring XRD1 beamline. Diffraction patterns have been collected in the temperature range 100 - 300 K. The STM and STS experiments have been performed on \emph{in situ}-cleaved samples in a UHV chamber  equipped with an Omicron Cryogenic STM, operating in constant-current mode at 10 K.  The STS measurements were recorded using a lock-in amplifier with a modulation voltage of 5 mV$_{rms}$ and a modulation frequency of 687 Hz. The bias voltage is given with respect to the sample.  \emph{Ab initio} calculations have been carried out, for the crystal lattice parameters taken from Ref. \cite{Fje_1986}, within the density functional theory (DFT) framework employing the generalized gradient approximation (GGA) as implemented in the QUANTUM ESPRESSO package \cite{Giannozzi_2009}. Spin-orbit effects were accounted for using fully relativistic norm-conserving pseudopotentials \cite{Corso_PRB_05}. 


\begin{figure}[t!]
  \centering
   \includegraphics[width = 0.45 \textwidth]{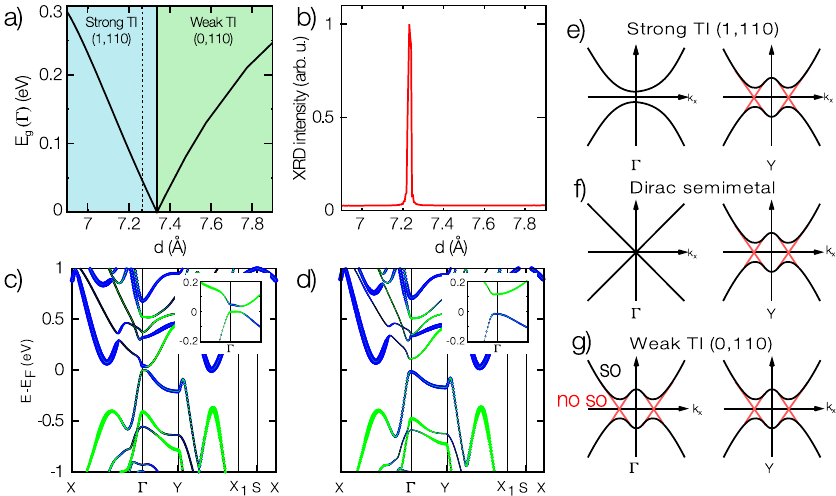}
  \caption[didascalia]{(a) Band gap at $\mathrm{\Gamma}$ as a function of the interlayer distance \emph{d}. (b) X-ray diffraction intensity measured at 300 K. (c, d) Calculated band structure of bulk $\mathrm{ZrTe_5}$ for the experimental interlayer distance \emph{b}/2 = 7.23 $\mathrm{\AA}$ and an enlarged interlayer distance of 7.5 $\mathrm{\AA}$, respectively. Green and blue colors indicate the weight of the states on the $\mathrm{Te_ d}$ and $\mathrm{Te_ z}$ $p$ orbitals. The inset shows a closeup in the bulk gap region at  $\mathrm{\Gamma}$.  (e - g) Schematic band diagram at $\mathrm{\Gamma}$ and $\mathrm{Y}$ in the $\mathrm{k_a}$ direction for the STI phase (e), the Dirac semimetal phase (f) and the WTI phase (g), with (black line) and without (red line) spin-orbit coupling.

   }
  \label{fig:ARPES_bands}
\end{figure}



\begin{figure*}[t]
 \includegraphics[width = 0.75 \textwidth]{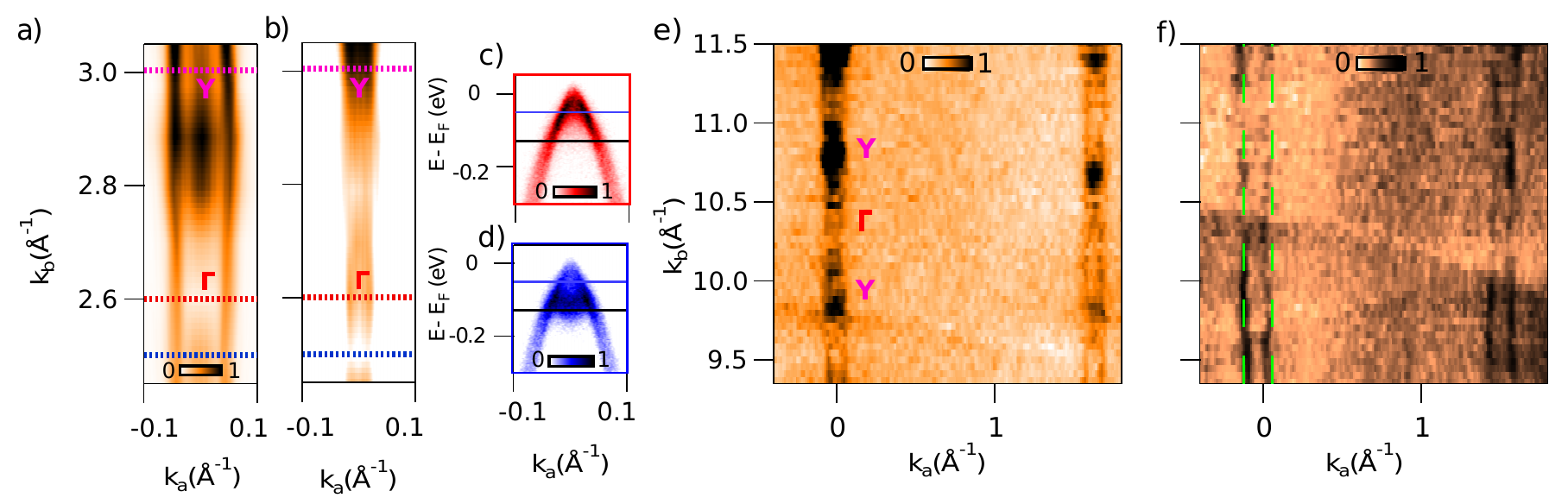}
  \caption[didascalia]{ Photon energy dependent study of $\mathrm{ZrTe_5}$. (a), (b) UV ARPES  constant energy map (CEM) in the ($\mathrm{k_a} $,\,$\mathrm{k_b}$) plane. The CEMs correspond to $-$130 and $-$50 meV, respectively. (c) and (d) ARPES images with black and blue lines indicating the energy of the CEM of panel (a) and (b), respectively. In (a), two contours are resolved: a 3D pocket centered at the $\mathrm{Y}$ points associated to BVB and 2D lines associated to SS. (e) and (f) CEM from the SX ARPES study with LV and LH polarization, respectively. (e) BVB contours with 3D bulk dispersion are visible, reproducing the results of (a). (f) Only open contours associated to SS are visible, thus indicating the different symmetries of BVB and SS. Green dashed lines indicate the 2D open contour of the surface state in (f).

   }
  \label{fig:Kz_CEM}
\end{figure*}



Figure 1 (a) shows the $\mathrm{ZrTe_5}$ orthorombic crystal structure, with $Cmcm$~(63) space group symmetry. The $\mathrm{ZrTe_5}$ natural cleavage plane results from two $\mathrm{ZrTe_3}$ chains oriented along the \emph{a} crystallographic direction. Each chain is formed by prisms with Zr (teal) and $\mathrm{Te_ a}$ and $\mathrm{Te_ d}$ (black) atoms at the apex and at the base of the prism, respectively. The chains are connected by $\mathrm{Te_ z}$ atoms along the \emph{c} axis, with \emph{a} = 3.99 $\mathrm{\AA}$ and \emph{c} = 13.73 $\mathrm{\AA}$, as determined by x-ray powder diffraction \cite{Fje_1986}. Each crystal cell contains two $\mathrm{ZrTe_5}$ planes piled along the \emph{b} axis. For this reason we indicate the interlayer distance with \emph{b}/2, whose value of 7.25 $\mathrm{\AA}$ reported in the literature \cite{Fje_1986} well matches our experimental finding, as it will be described later.

Figure 1 (b) and (c) show the UV ARPES measurements along the chain direction, $\mathrm{k_a}$, for two photon energies corresponding to the $\mathrm{\Gamma X}$ and $\mathrm{Y X_1}$ high symmetry directions of the 3D Brillouin zone (BZ), respectively. In Fig. 1 (b), VB disperses almost linearly at the $\mathrm{\Gamma}$ point, with a small charge carriers density at the Fermi level ($\mathrm{E_F}$).  In Fig. 1 (c), we observe additional spectral weight for $\mathrm{E - E_F = 0.1}$ eV, for wave-vectors smaller than the one of the linearly dispersing state, as indicated by a black arrow. The first state still disperses linearly across $\mathrm{E_F}$. The second state reaches its maximum below $\mathrm{E_F}$, and its dispersion deviates from linear, forming a \emph{M}-like shape. The dispersion of two distinct states is better conveyed by Fig. 1 (d). It shows a zoom at $\mathrm{E_F}$ for a different wave-vector orthogonal to the sample surface $\mathrm{k_b}$, whose value is indicated by color lines in Figure 3 (a) and (b).

Two states are clearly resolved at the top of VB. One has bulk character (BVB), as it displays a $\mathrm{k_b}$ dependence evolving from a linear dispersion at the $\mathrm{\Gamma}$ point to a \emph{M}-like shape at the $\mathrm{Y}$ point. The second state disperses linearly crossing $\mathrm{E_F}$ with no appreciable photon energy dependence. Hence, its  2D character is consistent with the behavior of a surface state (SS). The presence of two nearly degenerate states at the top of BVB has been reported both above and below $\mathrm{T^*}$ (see suppl. information \cite{Suppl_mat}).

These results shed new light on two recent transport studies \cite{Niu_arxiv_2015, pariari_arxiv_16}. The former shows two states with semiconducting and semi-metallic behaviors \cite{Niu_arxiv_2015}, the latter proposes the simultaneous existence of surface massless and bulk massive Dirac particles in $\mathrm{ZrTe_5}$  \cite{pariari_arxiv_16}.  

The measured band structure is well reproduced by \emph{ab initio} calculations, as shown in Figure 1 (e).  The projected bulk band structure (blue area) captures the evolution of BVB from linear to the \emph{M}-like shape. More importantly, a surface state (red line) disperses within the band gap, as enhanced in the inset, by zooming a small region across $\mathrm{E_F}$. 



When \emph{ad hoc} measured crystal structure parameters are used, DFT calculations predict that $\mathrm{ZrTe_5}$ is a STI. When the spin-orbit coupling is not taken into account, conduction and valence bands are degenerate along a nodal line in the $\mathrm{k_y=0}$ plane. Spin-orbit coupling breaks this degeneracy resulting in a STI phase with a band inversion at $\mathrm{Y}$. The product of parities of the valence band states at the $\mathrm{Y}$ point $k_Y=(\frac{1}{2},\frac{1}{2},0)$ is reversed with respect to the other time-reversal invariant momentum (TRIM) k-points which indicates that the $Z_2$ topological invariants are $(1,110)$.  Upon increasing the interlayer distance \emph{b}/2, the gap at $\mathrm{\Gamma}$ decreases, closes and reopens (see Fig.~2 (a) ) in a phase where the product of parities of the valence states at $\mathrm{\Gamma}$ is reversed and $\mathrm{ZrTe_5}$  is a WTI with invariants $(0,110)$. The observation of a topological phase transition is in agreement with the results in Ref.\cite{Weng_PRX_2014}. At the transition between the weak and strong phases, for \emph{b}/2\,=\,7.35\,$\mathrm{\AA}$, the gap at $\mathrm{\Gamma}$ is closed and $\mathrm{ZrTe_5}$ lies in a Dirac semimetal phase. However, it is of paramount importance to note that this Dirac semimetal phase requires a fine tuning of the interlayer distance. Figure 2 (b) shows the results of our x-ray diffraction study at 300 K. The experimental value of the interlayer distance \emph{b}/2\,= \,7.23\,$\mathrm{\pm}$ 0.02 $\mathrm{\AA}$. The interlayer distance monotonically decreases as the temperature is lowered in the range from 300 to 100 K (see suppl. information \cite{Suppl_mat}). This is consistent with a previous XRD study \cite{Fje_1986}, indicating that the $\mathrm{ZrTe_5}$ sample investigated in our work lies in the STI phase at temperatures both above and below $\mathrm{T^*}$. 

The band structure of $\mathrm{ZrTe_5}$ in the strong phase obtained from the experimental structure, and in the weak phase, as obtained for a strained structure with  \emph{b}/2 = 7.5 $\mathrm{\AA}$, are shown in Fig. 2 (c) and (d), respectively. The band inversion taking place at $\mathrm{\Gamma}$ during the topological phase transition can be highlighted by computing the weight of the states on the $p$ orbitals of the $\mathrm{Te_ d}$ and $\mathrm{Te_ z}$ atoms, indicated by blue and green colors in Fig. 2 (c - d), respectively. The three topological phases differ by their dispersion along the $\mathrm{k_a}$ direction at the TRIM points $\mathrm{\Gamma}$  and $\mathrm{Y}$ as illustrated by the schematic band diagrams in Fig. 2 (e - g), with (black lines) and without (red lines) spin-orbit coupling (SOC). While for each phase the dispersion around $\mathrm{Y}$ is  \emph{M}-shaped, the dispersion around $\mathrm{\Gamma}$ changes from parabolic in the strong topological phase to a  \emph{M}-shaped dispersion, reflecting the presence of a band inversion, in the weak topological phase.


In order to correctly determine the dimensionality of SS and BVB, we carried out a photon energy dependent study of the $\mathrm{ZrTe_5}$ band structure.  Figure 3 (a) and (b) shows two constant energy maps (CEMs) in the plane formed by the chain direction, $\mathrm{k_a}$, and the surface orthogonal, $\mathrm{k_b}$. Similar results for the ($\mathrm{k_a}$, $\mathrm{k_c}$) planes are discussed in the suppl. information \cite{Suppl_mat}. The CEMs are extracted at $-$130 and $-$50 meV, respectively. These values correspond to energies immediately below (a) and above (b) the maximum of BVB at the $\mathrm{Y}$ point, as shown by the black and blue lines traced on the ARPES image of Fig. 3 (c) and (d). In Fig. 3 (a) and (b) dashed colored lines indicate the positions of the band dispersion shown in Fig. 1.  In Figure 3 (a) we identify two contours: the first contour forms a closed pocket centered at the $\mathrm{Y}$ point, it arises from BVB and its $\mathrm{k_b}$ dependence clearly reflects the bulk character of the state. The second contour is open, it arises from SS and it shows negligible $\mathrm{k_b}$ dispersions. In Fig. 3 (b), for energies larger than the maximum of BVB at $\mathrm{Y}$, only non-dispersing open contours are resolved.

The longer photoelectron mean free path in SX ARPES provides a more accurate mapping of 3D periodicity of the BVB dispersion \cite{strocov_2003}, as shown in Fig. 3 (e), displaying the SX photon energy scan for LV polarization. In this dataset the binding energy of the CEM is $-$50 meV, which corresponds to the same energy position of panel (a) with respect of the band structure, owing to the temperature dependent energy shift of the band structure \cite{Manzoni_PRL_2015, Wu_arxiv_2016}, (see suppl. information \cite{Suppl_mat}). We resolve clearly the BVB state with 3D dispersion, with three replica along $\mathrm{k_b}$ well reproducing the data of Fig. 3 (a). 

Furthermore, it is interesting to notice the different origin of BVB and SS, which is ensured by their different response to the photon polarization. In fact, figure 3 (f) shows a photon energy scan performed under the same experimental conditions of Fig. 3 (e), but with LH polarization. The 3D bulk pockets associated to BVB are not visible anymore and only the open contours associated to SS are still present, indicated by the green dashed lines. Accordingly to the SX ARPES geometry, this effect can be interpreted as a consequence of the different symmetries of BVB and SS, being the former even and the latter odd, with respect to the $\mathrm{ZrTe_5}$ symmetry plane. The different matrix element effects between the UV and SX ARPES results reflect the two different experimental geometries (see suppl. information \cite{Suppl_mat}).


\begin{figure}[t!]
  \centering
   \includegraphics[width = 0.4 \textwidth]{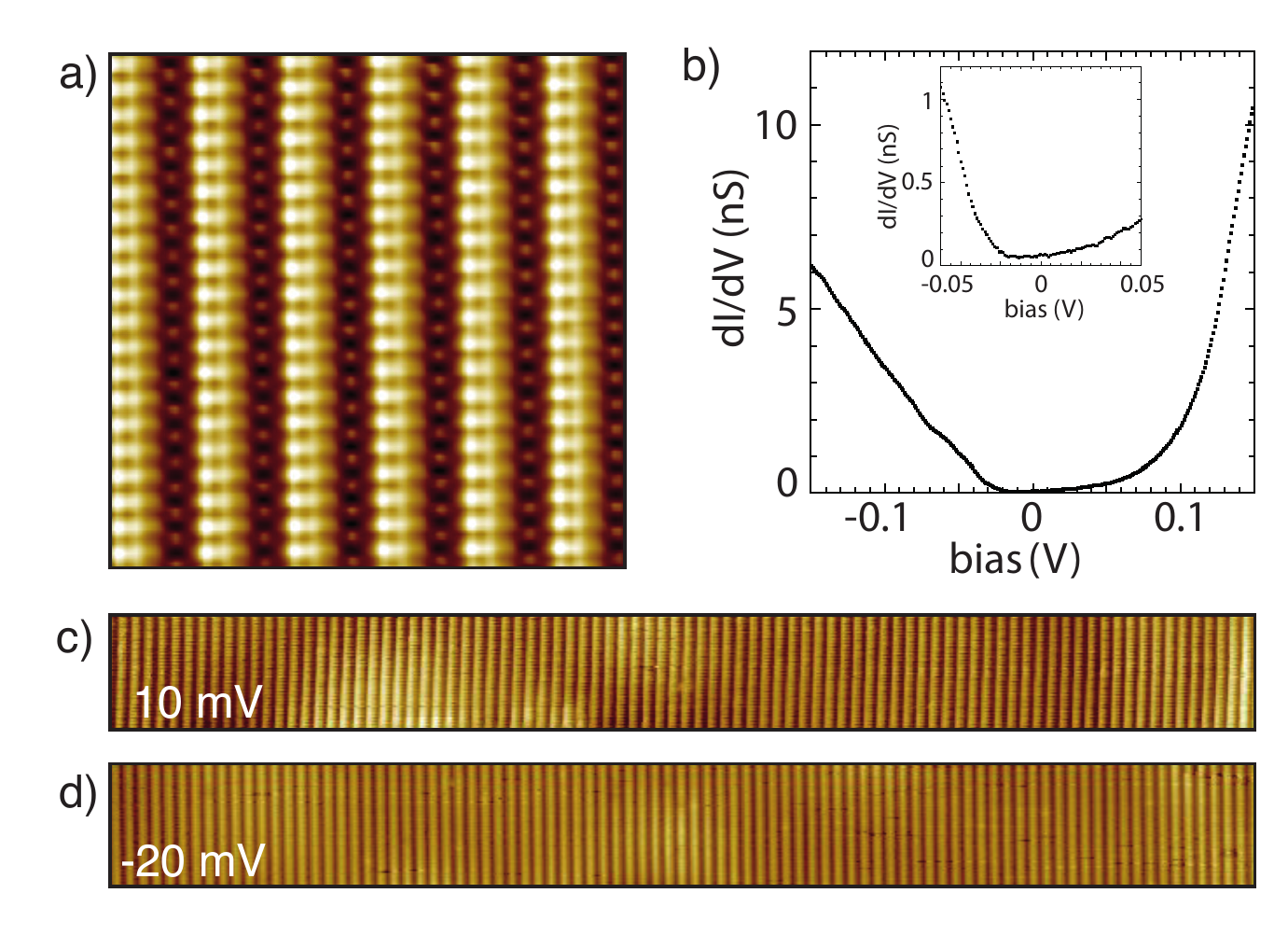}
  \caption[didascalia]{ (a) Atomic-resolution STM image of the $\mathrm{ZrTe_5}$ surface (8 nm $\mathrm{\times}$ 8 nm). Image conditions: V = -300 mV, I = 300 pA. (b) d$I$/d$V$ spectrum on the $\mathrm{ZrTe_5}$ surface. Set point: V = 0.15 V, I = 0.3 nA, V$_{mod}$ = 5 mV. Inset: detail of the same spectrum, plotted in the $\pm~50$ mV range, (c) and (d) STM topography image of the $\mathrm{ZrTe_5}$ surface (120 nm $\mathrm{\times}$ 13 nm) recorded at V = 10 mV and V = -20 mV respectively, I = 250 pA.

   }
  \label{fig:STM}
\end{figure}



We completed the occupied band structure mapping by performing STM/STS investigations. This allows to directly access both the occupied and unoccupied DOS, below and above $\mathrm{E_F}$. High-resolution STM images show the atomic structure of the $\mathrm{ZrTe_5}$ surface as chains of protruding pairs, separated by apparently monoatomic lines, as depicted in Fig. 4 (a). According to our ARPES results, the top of BVB lies in proximity of $\mathrm{E_F}$. Hence, in order to obtain insight about the conductivity in the bulk band gap, we performed STS in a low bias voltage range. Fig.\,4\,(b) presents the d$I$/d$V$ spectrum in the $\pm~150$ meV range. The negative energy values correspond to the occupied states. Within this range, the curve appears to be characterized by a monotonic decrease/increase of conductivity with a minimum at around $-$15 mV, as clearly seen in the inset of the Fig. 4 (b), which represents a plot of the same spectrum in the $\pm~50$ mV range.

Remarkably, within our experimental resolution, the conductivity of the sample never vanishes. To further elaborate on the latter point, we performed STM imaging at very low voltage bias, as depicted in Fig. 4 (c) and (d). Both below ($-$20 mV) and above ($+$10 mV) $\mathrm{E_F}$, we are able to image the atomic stripes of the $\mathrm{ZrTe_5}$ structure, even over areas which are larger than 100 nm. Furthermore, the observation of standing waves in the d$I$/d$V$ conductance maps in proximity of step regions (see the suppl. information \cite{Suppl_mat}) suggests that the metallic DOS at  $\mathrm{E_F}$ arises from a two-dimensional surface state.

In conclusion, in this Letter we report a comprehensive experimental and theoretical investigation of the electronic and structural properties of the bulk $\mathrm{ZrTe_5}$. Altogether, we show that $\mathrm{ZrTe_5}$ is a bulk STI both above and below the temperature of the resistivity peak T*. It lies near a topological phase transition to a WTI phase that can be triggered by an increase of the interlayer distance. This observation reconciles the recent conflicting results about the topological character of $\mathrm{ZrTe_5}$ \cite{Valla_arx_2015, Zhou_arx_2015, Chen_PRL_2015, Chen_PRB_2015, Manzoni_PRL_2015, Li_arxiv_2016, Wu_arxiv_2016, Niu_arxiv_2015, pariari_arxiv_16}. Importantly, $\mathrm{ZrTe_5}$ is not a 3D Dirac semimetal, as this phase is not protected by crystalline symmetry and can only be realized precisely at the topological phase transition. Nonetheless, the band dispersion is close to a 3D Dirac semimetal thus explaining the recent optical and electron transport measurements \cite{Valla_arx_2015, Zhou_arx_2015, Chen_PRL_2015, Chen_PRB_2015, Niu_arxiv_2015, pariari_arxiv_16}. The close vicinity to the topological phase transition suggests the possibility of engineering the topological phase by a small variation of the interlayer distance, entering in the WTI phase. This can be achieved by changing the concentration of defects, that may vary depending on the growth conditions, thus also accounting for the experimental observation of different T* in different studies  \cite{Valla_arx_2015, Manzoni_PRL_2015, Yu_Pan_arxiv_16, Sabongi_1980} . The topological phase transition may also be achieved in a controlled way either via alkali metal intercalation, as exploited for decoupling  $\mathrm{MoS_2}$ single layer \cite{Meevasana_NL_2014}, or via chemical substitution. This will require further ARPES investigations of \emph{ad hoc} grown samples with controlled doping and/or defect density. Finally, the observed monotonic decrease of \emph{b}/2 as a function of temperature suggests that, once the WTI phase is achieved, transition to the STI phase can be controlled by temperature, thus making $\mathrm{ZrTe_5}$ a versatile platform for spintronics applications.



We gratefully acknowledge fruitful discussions with Marco Grioni and Doriano Lamba. This work was supported in part by the Italian Ministry of University and Research under Grant Nos. FIRBRBAP045JF2 and FIRB-RBAP06AWK3 and by the European Community Research Infrastructure Action under the FP6 "Structuring the European Research Area" Program through the Integrated Infrastructure Initiative "Integrating Activity on Synchrotron and Free Electron Laser Science", Contract No. RII3-CT-2004-506008. G.A. and O.V.Y. acknowledge support by the NCCR Marvel and the ERC Starting grant ``TopoMat'' (Grant No. 306504). First-principles calculations have been performed at the Swiss National Supercomputing Centre (CSCS) under project s515. F. B.  acknowledge support by the SNF through the project $\mathrm{200021\_146890}$. 






\end{document}